\documentstyle[preprint,prl,aps,epsf]{revtex}
\title{Correlations in Transmission of Light through a Disordered
Amplifying Medium}
\author{A. A. Burkov, A. Yu. Zyuzin}
\date{\small A.F.Ioffe Physical-Technical Institute,\\
194021 Saint-Petersburg, Russia}
\begin{document}
\maketitle
\begin{abstract}
The angular and frequency correlation functions of the transmission
coefficient for light propagation through a strongly scattering amplifying
medium are considered.
It is found that just as in the case of an elastic scattering medium the
correlation function consists of three terms.
However, the structure of the terms is rather different.
Angular correlation has a power-law decay and exhibits oscillations.
There is no "memory effect" as in the case of an elastic medium.
Interaction between diffusion modes is strongly enhanced near the lasing
threshold.
Frequency correlation scale decreases close to the lasing threshold.

We also consider time correlations of the transmission in
the case of nonstationary inhomogeneities.
We find short- and long-range time correlations.
The scale of the short-range correlation decreases, while the long-range
correlation scale becomes infinite near the threshold.
\end{abstract}
\newpage
\section{Introduction}

In recent years novel findings have been uncovered in speckle patterns of
waves coherently transmitted through a disordered medium.
The most interesting among them are intensity fluctuations with long-range
correlation including power-law and "infinite-range" correlations.
The latter are related to the "universal conductance fluctuations" in
disordered metals.
The theory of speckle arising due to the multiple scattering was
developed in~\cite{Stephen87,Zyuzin87,Feng88,Shapiro89}.
Experimental testing was done for microwave~\cite{Genack89} as well
as for visible light~\cite{Lagendijk90}.

Recently the study of light propagation in disordered
amplifying media has begun.
This problem is different from a conventional
radiation diffusion in a disordered medium because here scattering also
provides a feedback for light generation and, therefore,upon a certain
condition an amplifying disordered medium becomes
a generator of coherent radiation~\cite{Letokhov67}.
In~\cite{Lawandy94} authors observed laser action in a dye dispersed in
a weakly scattering medium.
Amplification allows for probing the longest paths that light travels in
the system.
This brings a modification of coherent backscattering~\cite{Zyuzin94}
which was observed in~\cite{Wiersma95}.
It would be natural to carry on these studies by an observation of
speckle correlations of light scattered from an amplifying medium.

In this paper we consider correlations  in intensity of light transmitted
through a disordered amplifying medium.
Some statistical properties of such a medium were discussed
in~\cite{Zyuzin95,Pradhan94}.
Statistical properties of lasing states~\cite{Zyuzin95} and transmission
distribution function~\cite{Pradhan94} were considered.

Here we calculate the frequency and angular correlation functions
of the transmission coefficient of a diffusive amplifying medium.
We approach the problem in a way that resembles that of~\cite{Zyuzin87,Feng88}.

We show that near the lasing threshold the correlation function consists
of three terms just as in the case of an elastic medium.
However, the structure of the terms is different.
The first term ( $C^{(1)}$ ) which describes the contribution of
noninteracting diffusion modes
has a power-law decay . There is no "memory effect":
the term depends separately on the momentum differences of the incoming
and outgoing waves.

The contributions from interaction of diffusion modes are more singular
near the lasing threshold compared to $C^{(1)}$.
The $C^{(2)}$ term has the same dependence on the incoming and outgoing
momentum differences as the $C^{(1)}$.
The "infinite-range" correlation observed in an elastic medium
is also present in our case, but it has a much larger relative magnitude.

In frequency dependence of the correlation function
near the threshold there
appears a correlation scale which decreases close to the lasing threshold.
The dependence is the same in all three terms.

We also calculated time correlation function
in the case of a nonstationary dielectric function fluctuations,
caused by freely moving point scatterers.
We obtain short- and long-range time correlations.

Our paper is organized as follows.
In section 2 we introduce the basic quantities which we use in our
calculations such as the average Green's function of the wave equation and
the field-field correlation function.
In section 3 we define the                                      
transmission coefficient and calculate the average value of it.
We also discuss here the weak localization correction to the
diffusion coefficient.
In section 4 we calculate the angular and frequency correlation
functions of the transmission coefficient and time correlation
in the case
of nonstationary inhomogeneities.

\section{Field-field correlation function.}

We consider the propagation of electromagnetic waves in a weakly disordered
amplifying medium.
We start with a scalar linear wave equation
derived from Maxwell`s equations:
\begin{equation}
[\,\nabla^2\,+\,k^2\epsilon( \vec r )\,]\,E\,=\,0
\end{equation}
Here $k\,=\,\frac{2\pi}{\lambda}$ is the wavenumber, $\epsilon(\vec r)$ ---
the complex dielectric function.
It is of the form $ \epsilon\,=\,\epsilon'\,+\,i\,\epsilon''$, where
the average part:
\begin{equation}
\epsilon'(\omega)\,=\,1\,-\,\frac{2\pi(N_L-N_U)e^2}{m\omega_0}
\frac{\omega-\omega_0}{(\omega-\omega_0)^2+\gamma^2}
\end{equation}
\begin{equation}
\epsilon''(\omega)\,=\,\frac{2\pi(N_L-N_U)e^2}{m\omega_0}
\frac{\gamma}{(\omega-\omega_0)^2+\gamma^2}
\end{equation}
describes the response of oscillating dipoles of frequency
$\omega_0$ and mass $m$.
$N_U$ and $N_L$ are the densities of the dipoles in the excited and low
states correspondingly ( $N_U > N_L$ ).
$\gamma^{-1}$ corresponds to the lifetime of the excited state.

We will consider the propagation of electromagnetic waves with frequencies
$\omega$ in an interval near the atomic frequency much smaller than broadening
$\gamma$ so that we can neglect the dispersion of the dielectric function.
Thus we consider $\epsilon''$ as a negative constant and suppose that
$\epsilon'\,\approx\,1$.

Now let some defects or impurities cause fluctuations of the real part
of the dielectric function.
Then:
\begin{equation}
\epsilon'(\vec r)\,=\,1\,+\,\delta\epsilon(\vec r)
\end{equation}
where $\delta\epsilon(\vec r)$ is a random function with the following
properties:
$$\langle\delta\epsilon(\vec r)\rangle\,=\,0$$
and
\begin{equation}
k^4\langle\delta\epsilon(\vec r)\delta\epsilon(\vec r\,')\rangle\,=\,\Lambda\,
\delta(\vec r\,-\,\vec r\,')
\end{equation}
Here angular brackets denote averaging over realizations of the
disorder.

$\Lambda$ is simply related to the mean free path of the radiation $l_0$:
\begin{equation}
\Lambda\,=\,\frac{4\pi}{l_0}
\end{equation}

In the lowest order of the disorder parameter $(kl_0)^{-1}\,\ll\,1$
the average Green's function
of the equation~(1) is:
\begin{equation}
D(r)\,=\,-\,\frac{1}{4\pi r}\,\exp\left(ikr-\frac{r}{2l}\right)
\end{equation}
where $\frac{1}{l}\,=\,\frac{1}{l_0}\,-\,k|\epsilon''|\,=\,
\frac{1}{l_0}\,-\,\frac{1}{l_{amp}}$.

The length $l_{amp}\,=\,\frac{1}{k|\epsilon''|}$ is the amplification
length.

The properties of light energy transport in a disordered medium are
determined by the field-field correlation function:
\begin{equation}
\Gamma_{\omega\omega'}(\vec r_1,\vec r_2)\,=\,\langle E_{\omega}
(\vec r_1)\,E_{\omega'}^*(\vec r_2)\rangle
\end{equation}
Here $E_{\omega}$ is the field of frequency $\omega$.

$\Gamma_{\omega\omega}(\vec r,\vec r)$ equals $\langle I_{\omega}(\vec r)
\rangle$ --- the average intensity at a point $\vec r$ of the field of
frequency $\omega$.

In the leading order of the disorder parameter the field-field correlation
function is given by the sum of the ladder diagrams:
\begin{equation}
\Gamma_{\omega\omega'}(\vec r_1,\vec r_2)\,=\,\frac{4\pi}{l_0}\int d\vec r\,'
d\vec r\,''D( \vec r_1\,-\,\vec r\,')
D^*( \vec r_2\,-\,\vec r\,')\,P_{\omega\omega'}
(\vec r\,',\vec r\,'') E_{n\omega}(\vec r\,'') E^*_{n\omega'} (\vec r\,'')
\end{equation}
Here
$E_{n\omega}(\vec r)$ is the nonscattered field at a point $\vec r$
and $P_{\omega\omega'}$ satisfies the equation:
\begin{equation}
\left(D\nabla^2\,+\,\frac{1}{\tau_{amp}}\,+\,i\,\,\delta\omega\right)
P_{\omega\omega'}(\vec r)\,=\,-\,\frac{1}{\tau}\,\,\delta(\vec r)
\end{equation}
where $\delta\omega\,=\,\omega\,-\,\omega'\,,\,\tau_{amp}\,=\,
l_{amp}/c$, $\tau\,=\,l/c$ and $D\,=\,lc/3$ is a diffusion coefficient.

Let us also introduce the light energy current density:
\begin{equation}
\vec J_{\omega}(\vec r)\,=\,
\frac{ic}{2k}(E_{\omega}\vec\nabla E^*_{\omega}\,-\,E^*
_{\omega}\vec\nabla E_{\omega})
\end{equation}
It can be easily shown using (9) (see \cite{Shapiro89}) that:
\begin{equation}
\langle \vec J_{\omega}(\vec r)\rangle\,=\,
-\,D\vec\nabla\langle I_{\omega}(\vec r,t)\rangle
\end{equation}
Taking into account a conservation law:
\begin{equation}
\frac{\partial \langle I_{\omega}(\vec r, t) \rangle}{\partial t}\,+\,
\vec\nabla \langle\vec J_{\omega}(\vec r)\rangle\,=\,
\frac{\langle I_{\omega}(\vec r)\rangle}
{\tau_{amp}}
\end{equation}
one obtains a diffusion equation for the average intensity:
\begin{equation}
D\nabla^2\langle I_{\omega}(\vec r,t)\rangle\,+\,
\frac{\langle I_{\omega}(\vec r,t)\rangle}
{\tau_{amp}}\,=\,\frac{\partial \langle I_{\omega}(\vec r, t)\rangle}{\partial t}
\end{equation}

Equations~(12)---(14) describe the propagation of radiation in a medium
with dimensions which are larger than the mean free path~$l_0$.
We consider a sample in the form of a slab with dimensions $L_x,\,L_y\,$
and $L_z$, where the sides $z\,=\,0,L_z$ are open and the other side surfaces
are perfectly reflecting.
We suppose that $L_x \sim L_y \sim L_z$.
The slab has a critical width $L_{cr}\,=\,\pi\,\sqrt{D\tau_{amp}}$.
If $L_z \geq L_{cr}$ the system becomes a multimode generator
\cite{Letokhov67,Zyuzin95}.
We will measure the size of the system in the units of the critical
size:
\begin{equation}
L_z\,=\,L_{cr}(1\,-\,\Delta)
\end{equation}
We suppose that $\Delta \ll 1$.

The case of $L_z\gg l_0$ corresponds to the experimental setup of~\cite{Wiersma95}.
The opposite limit $L_z\ll l_0, L_x, L_y$ was realized in~\cite{Lawandy94}.
Critical conditions for this case are very different from~(15) and
the critical value of amplification length is of the order of the mean
free path~\cite{Zyuzin95a}.
The total phase diagram amplifier-generator is shown on fig.1.
More discussion of the critical conditions can be found in~\cite{Burkov96}.
Calculations in the case $L_z\leq l_0$ are beyond the scope of our paper.

\section{The average transmission coefficient}

Consider the transmission of monochromatic light through
the slab described in the previous section.

We define the transmission coefficient of the light of frequency
$\omega$ incident at an angle $a$ on the plane $z\,=\,0$ and
transmitted at an angle $b$ as the ratio of the emergent flux per unit
solid angle around the direction $\vec q_{b}$ to the incident flux.

From (9) and (11) one has for the transmission coefficient:
\begin{equation}
\langle T_{ab\omega}\rangle\,=\,\frac{-l c\, \cos^2 b}{cI_0L_xL_y}
\int d^2 r\, \frac{\partial}{\partial z}\,
\langle I_{\omega}(\vec r)\rangle_{z=L_z}
\,=\,\frac{-\,l^3\,\cos^2 b}{L_x L_y}\,
\int d^2 r\,d^2 r'\,\frac{\partial}{\partial z}
\,\frac{\partial}{\partial z'}\,P_{\omega\omega}
(\vec r, \vec r\,')_{z=L_z,z'=0}
\end{equation}
where $I_0$ is the incident intensity.

The corresponding to the equation (16) diagram is shown on fig.2.

For the equation (10) for function $P_{\omega\omega'}(\vec r\,,\vec r\,')$
we assume the following boundary conditions: $P_{\omega\omega'}$ is zero
on the boundaries $z\,=\,0,L_z$ and the current through the side surfaces
is zero.
Then near the threshold one can calculate the function $P_{\omega
\omega'}(\vec r\,,\vec r\,')$ as:
\begin{equation}
P_{\omega\omega'} ( \vec r\,,\vec r\,') \,=\,\frac{3L_{cr}}
{l^2{\pi}^2L_xL_y} \frac{sin(\frac{\pi}{L_{cr}} z) sin(\frac{\pi}{L_{cr}}
z')}{\Delta\left(1\,-\,i\frac{\delta\omega}{\Omega} \right)}
\end{equation}
where $\Omega\,=\,2\Delta/\tau_{amp}$.
In (17) we neglect contributions which are nonsingular
at $\Delta \rightarrow 0$.
Hence one obtains:
\begin{equation}
\langle T_{ab\omega}\rangle\,=\,\frac{3l}{L_{cr}\Delta}  \cos^2b
\end{equation}
Integration over angles gives the average coefficient of the total
transmission:
\begin{equation}
\langle T \rangle\,=\,\frac{l}{L_{cr}\Delta}
\end{equation}
The transmission coefficient contains an additional amplification factor
$\Delta^{-1}$ compared to the case of an elastic medium.

Let us now consider the weak localization correction to the diffusion
coefficient in equation (14) in the limit $\Delta\rightarrow 0$.
Weak localization has been extensively studied in physics of
disordered conductors (~see for review~\cite{Altshuler83}~)
and we may use the well
known results.
Localization correction to the diffusion constant is given by maximally
crossed diagrams.
The analytical expression is:
\begin{equation}
\tilde {D}(\delta\omega)\,=\,D \left( 1\,-\,\frac{2\pi l}{k^2 V}
\int d\vec r\, C_{\omega\omega'}(\vec r,\vec r)\right)
\end{equation}
Here $V\,=\,L_xL_yL_z$ is the volume of the slab and cooperon
$ C_{\omega\omega'}(\vec r,\vec r\,')$ satisfies in the case of stationary
inhomogeneities eq.(10) and has therefore form~(17).
Substituting it in (20) one obtains:
\begin{equation}
\tilde {D}(\delta\omega)\,=\,D\left[ 1\,-\,\frac{3}{\pi \Delta
\left(1\,-\,i
\frac{\delta\omega}{\Omega} \right) g} \right]
\end{equation}
Here $g\,=\,\frac{k^2L_xL_yl}{L_{cr}}\,\gg\,1$ is a quantity analogous to the
dimensionless conductance of a weakly disordered metal.
Let us note that there is no correction to $1/\tau_{amp}$ in eq.(14).
For $\Delta\,\rightarrow\,0$ the correction in (21) coincides with that
for electrons in a small closed metallic particle.

For small $\delta\omega$, such as $\delta\omega\,\sim\,
\frac{1}{g\tau_{amp}}\,\sim\,
\frac{\omega}{k^3V}$, the correction is of order unity.
In our case it means that at a time $\sim (\delta\omega)^{-1}$
diffusion approach breaks down and transport is dominated by
only a few lasing states.
At zero $\delta\omega$ the condition of smallness of the correction is
$g\Delta\ >1$.

\section{Intensity-intensity correlations}

Average quantities do not describe the transmission of light in a random
system completely.
In this section we consider fluctuations of the transmission.
We study the following correlation function:
\begin{equation}
C_{aba'b'}^{\omega\omega'}\,=\,\langle \delta T_{ab\omega} \delta T_{a'b'
\omega'} \rangle
\end{equation}
where $ \delta T_{ab\omega}\,=\,T_{ab\omega}\,-\,\langle T_{ab\omega}
\rangle$.

It might be represented as \cite{Feng88}:
\begin{equation}
C_{aba'b'}^{\omega\omega'}\,=\,C^{(1)}\,+\,C^{(2)}\,+\,C^{(3)}
\end{equation}
Contribution $C^{(1)}$ describes noninteracting diffusion modes.
Interaction between them brings contributions $C^{(2)}$ and
$C^{(3)}$.
The diagram corresponding to $C^{(1)}$ is shown on fig.3.
The calculations leading to an expression for $ C^{(1)} $ are
analogous to the calculations of the average transmission coefficient.
The only modification needed is to take into account a change in the phase
of the incident or transmitted wave due to a change in the angle of
incidence or transmission:
\begin{equation}
C^{(1)}\,=\,\frac{\,l^6\,\cos^2 b\,\cos^2 b'}{L_x^2 L_y^2}\,
\left|\int d^2 r\,d^2 r'\,\exp\left(-i\delta\vec q_b\vec r\right)
\,\exp\left(i\delta\vec q_a\vec r\,'\right)\,
\frac{\partial}{\partial z}
\,\frac{\partial}{\partial z'}\,P_{\omega\omega'}
(\vec r, \vec r\,')_{z=L_z,z'=0} \right|^2
\end{equation}
Here
$ \delta\vec q_{a,b}$ is the difference between the wave vectors
of the incident($a$) or transmitted($b$)
waves corresponding to different measurements.

From (17) and (24) one obtains:
\begin{equation}
C^{(1)} \,=\,\langle T_{ab\omega} \rangle \langle T_{a'b'\omega'}
\rangle F(\delta\vec q_a) F( \delta\vec q_b) \frac{1}
{ 1\,+\,\left(\frac{\delta\omega}{\Omega}\right)^2}
\end{equation}
where
$$F(\delta\vec q)\,=\,\frac{sin^2\left( \delta q_{x}\frac{L_x}{2} \right) }
{ \left( \delta q_{x}\frac{L_x}{2} \right)^2}\, \frac{sin^2\left( \delta
q_{y} \frac{L_y}{2} \right) } { \left( \delta q_{y} \frac{L_y}{2}
\right)^2}$$
$C^{(2)}$ term can be represented as
$C^{(2)}\,=\,C^{(2)} (\delta\vec q_a)\,+\,C^{(2)} (\delta\vec q_b) $.
Due to reciprocity relations both terms are equal for $ \delta\vec q_a\,=\,
\delta\vec q_b$.

Let us note the following.
The scatterers in the disordered medium can be considered as sources
of light, which are phase or amplitude correlated.
Accordingly we can divide the terms constituting the speckle correlation
function in two groups.
Terms $C^{(1)}$ and $C^{(2)} (\delta\vec q_b)$ correspond to phase correlated,
$C^{(2)} (\delta\vec q_a)$ and $C^{(3)}$ --- to amplitude correlated
sources.

To calculate the amplitude correlated terms it is convenient to use
the Langevin approach~\cite{Zyuzin87}.
According to this approach the fluctuating part
$\delta\vec J_{\omega}\,=\,\vec J_{\omega}\,-\,\langle \vec J_{\omega}
\rangle$
of the light energy
current can be expressed as:
\begin{equation}
\delta\vec J_{\omega}(\vec r)\,=
\,-\,D\vec {\nabla}\delta I_{\omega}(\vec r)\,+\,
\vec j_{ext\,\omega}(\vec r)
\end{equation}
where $\vec j_{ext\,\omega}(\vec r)$ is the Langevin random current.
Using conservation law:
\begin{equation}
\vec{\nabla}\delta\vec J_{\omega}(\vec r)\,=
\,\frac{\delta I_{\omega}(\vec r)}{\tau_{amp}}
\end{equation}
one can write for the fluctuating part of the intensity:
\begin{equation}
\left(D\,\nabla^2\,+\,\frac{1}{\tau_{amp}} \right)
\delta I_{\omega}(\vec r)\,=\,
\vec {\nabla} \vec j_{ext\,\omega}(\vec r)
\end{equation}
With the help of (26) and (28) we can obtain an expression for
the fluctuating part of the total current through the surface $z\,=\,L_z$.
To calculate the correlation function of the total current we need an
expression for the correlation function of the Langevin current.

Near the lasing threshold the Langevin currents correlation functions
are given by the following expressions:
\begin{equation}
\langle j_{ext\,\omega}^{\,i}(\vec r_1)
j_{ext\,\omega'}^{\,j}(\vec r_2) \rangle_1\,=
\,\frac{2\pi lc^2}{3k^2} \langle I_{\omega}(\vec r_1) \rangle^2
F(\delta\vec q_a)
\frac{1}{1\,+\,\left(\frac{\delta\omega}{\Omega}\right)^2}
\delta_{ij} \delta\,(\vec r_1\,-\,\vec r_2)
\end{equation}
\begin{equation}
\langle j_{ext\,\omega}^{\,i}(\vec r_1)
j_{ext\,\omega'}^{\,j}(\vec r_2) \rangle_2\,=
\,\left(
\frac{2\pi l^2 c}{3k^2}\right)^2  ( K_1\,+\,K_2\,+\,K_3)
\end{equation}
\begin{equation}
K_1\,=\,\delta_{ij}\,\delta\,(\vec r_1\,-\,
\vec r_2) \int d\vec r_3 |P_{\omega\omega'}(\vec r_1,\vec r_3)|^2
\vec {\nabla} \langle I_{\omega}(\vec r_3,a)\rangle\vec {\nabla}\langle
I_{\omega'}(\vec r_3,a')\rangle
\end{equation}
\begin{equation}
K_2\,=\,|P_{\omega\omega'}(\vec r_1,
\vec r_2)|^2\nabla_i\langle I_{\omega'}(\vec R_1,a')\rangle\nabla_j
\langle I_{\omega}(\vec r_2,
a)\rangle
\end{equation}
\begin{equation}
K_3\,=\,
|P_{\omega\omega'}(\vec r_1,\vec r_2)|^2
\nabla_i\langle I_{\omega}(\vec r_1,a)\rangle\nabla_j
\langle I_{\omega'}(\vec r_2,a')\rangle
\end{equation}
Correlation functions (29) and (30) contribute to
$C^{(2)} (\delta\vec q_a)$ and $C^{(3)}$ correspondingly.
Diagram for correlation function (29) is shown on fig.4
and diagrams for the three terms in
(30)---on fig.'s 5a,b,c~\cite{Zyuzin87}.

Calculating fluctuations of the total current we obtain:
\begin{equation}
C^{(2)}\,=\,C^{(2)} (\delta\vec q_a)\,+\,C^{(2)} (\delta\vec q_b)\,=\,
\frac{3}{4\pi g\Delta^2}\langle T_{ab\omega}\rangle
\langle T_{a'b'\omega'}\rangle\frac{1}{1\,+\,
\left(\frac{\delta\omega}{\Omega}\right)^2}[ F(\delta\vec q_a)\,+\,
F(\delta\vec q_b)]
\end{equation}
\begin{equation}
C^{(3)}\,=\,\frac{27}{16\pi^2g^2\Delta^4}\langle T_{ab\omega}\rangle
\langle T_{a'b'\omega'}\rangle\frac{1}
{1\,+\,\left(\frac{\delta\omega}{\Omega}\right)^2}
\end{equation}

Let us now discuss our main results (25), (34) and~(35).
Note first, that there is no "memory effect" in the $C^{(1)}$ term ---
it depends separately on $\delta\vec q_a$ and $\delta\vec q_b$.
The form-factor function $F(\delta\vec q)$ has an oscillating
power-law decay .

The form-factor is exactly the same as in the case of transmission through
an elastic scattering medium in the form of a long tube~\cite{Berkovits91}.
The "memory effect" is also absent in such a medium.

Integrating $C^{(2)}$ over the angles of transmission we obtain the mean
square fluctuation of the total transmission~\cite{Zyuzin95}:
\begin{equation}
\langle\,(\delta T)^2\,\rangle\,=\,\frac{3}{4\pi g\Delta^2}\,
\langle T \rangle^2
\end{equation}
One sees that the transmission near the lasing threshold
is not a self-averaging quantity: fluctuations of total transmission grow
faster than the average value.

The $ C^{(3)}$ term just as in the case of an elastic medium represents
an "infinite-range" correlation in momentum space.
However, the relative magnitude of this correlation
is much larger in our case due to the factor $\Delta^{-4}$.

Close to the threshold all terms in the speckle correlation function
have the same frequency dependence.
The scale of the frequency correlation is determined by a parameter
$\Omega\,=\,2 \Delta/\tau_{amp}$, which goes to zero at
$\Delta \rightarrow 0$.

Interaction between diffusion modes according to (34) and (35) is described
by a parameter $1/g \Delta^2$, contrary to the case of a nonamplifying
medium where the parameter is $1/g$.
The increase of interaction can be interpreted as follows.
In the case of an elastic disordered system transmission probes the path
of typical length of order $L_z^2/l$.
In our case amplification compensates for escape of a wave, therefore
transmission probes paths of much longer lengths and  probability of interference
increases.

We should point out that the value of parameter~$\Delta$ in our
formulas can not be infinitely small.
$\Delta\,=\,0$~corresponds to the threshold in the equation for the average
intensity.
Actually the threshold value of $\Delta$ is a random function of frequency,
such that its mean square fluctuation is of the order
of~$g^{-1}$~\cite{Zyuzin95}.
It means that our consideration is valid
for $g \Delta^2 \geq 1$.
Note that this condition is more restrictive than the condition of smallness
of the weak localization correction $ g \Delta > 1$ in eg.~(14).

So far we assumed that the fluctuations of the dielectric function
which scatter light are stationary.
However it is also interesting to consider the correlation
function $ C_{abab'}(t)\,=\,\langle \delta T_{ab}(t)
\delta T_{ab'}(0) \rangle$ ( $T_{ab}(t)$ is the transmission coefficient
$T_{ab}$ measured at time $t$)
in the case of nonstationary fluctuations.
Note that we will consider correlations with a fixed incident channel
$ (a\,=\,a')$.

In the simple case of freely moving point scatterers
\cite{Golubentsev84,Stephen88} one obtains:
\begin{equation}
C^{(1)}(t)\,=\,\langle T_{ab}\rangle \langle T_{ab'}\rangle\,
F(\delta \vec q_b)\,F_1(t)
\end{equation}
\begin{equation}
C^{(2)}(t)\,=\, \frac{3}{4\pi g\Delta^2}\, \langle T_{ab} \rangle
\langle T_{ab'} \rangle\,\left[ F_2(t)\,+\,F(\delta\vec q_b)\,F_3(t) \right]
\end{equation}
Here
$$ F_1(t)\,=\,\frac{1}
{\left( 1\,+\,\frac{t^2}{2\,\tau_{\lambda}^2\,\Omega\,\tau}\right)^2}$$
$$ F_2(t)\,=\,\Omega^2
\,\int_0^{\infty}
\,dt_1\,dt_2\,\exp\left\{ - \Omega\,
\left( t_1\,+\,t_2 \right) \right\}\,F_1(t_1-t_2+t)$$

$$ F_3(t)\,=\,\Omega^2
\,\int_0^{\infty}\,dt_1\,dt_2\,\exp\left\{ - \Omega
\left(1\,+\,\frac{t^2}
{2\,\tau_{\lambda}^2\,\Omega\,\tau}
\right)\,\left( t_1\,+\,t_2 \right) \right\}
\,F_1(t_1-t_2)$$

and $ \tau_{\lambda}$
is the time for a scatterer to move a wavelength.
We do not consider the $C^{(3)}$ term because
in the present case it does not
contain additional features compared to $C^{(1)}$ and $C^{(2)}$.

To derive the above results one needs to modify the dielectric function
fluctuation's correlation function (5) to take
into account the motion of the
scatterers.
In the case of freely moving scatterers with a characteristic velocity
much less than the velocity of light in the medium
this correlation function averaged over the
angle of scattering becomes~\cite{Golubentsev84,Stephen88}:
\begin{equation}
k^4\langle\delta\epsilon(\vec r, t)\delta\epsilon(\vec r\,', 0)\rangle\,=\,
\Lambda\, f(t)\,
\delta(\vec r\,-\,\vec r\,')
\end{equation}
where $ f(t)\,=\,(\tau_{\lambda}^2/t^2)\,\left[1\,-\,\exp(-t^2/
\tau_{\lambda}^2)\right]\,\approx\,1\,-\,t^2/2\tau_{\lambda}^2$
and $t$ is the time difference variable.

Then it is easy to  show that function $P_{\omega\omega'}(t)$
describing the correlation of fields
$ \left\langle E_{\omega}(t) E_{\omega'}^*(0) \right\rangle$
measured with a time
interval $t$ satisfies the equation:
\begin{equation}
\left(D\nabla^2\,+\,\frac{1}{\tau_{amp}}\,+\,i\,\,\delta\omega
\,-\,\frac{t^2}{2 \tau \tau_{\lambda}^2}\right)
P_{\omega\omega'}(\vec r, t)\,=\,-\,\frac{1}{\tau}\,\,\delta(\vec r)
\end{equation}
Therefore near the threshold it is given by an expression:
\begin{equation}
P_{\omega\omega'} ( \vec r\,,\vec r\,', t) \,=\,\frac{3L_{cr}}
{l^2{\pi}^2L_xL_y} \frac{sin(\frac{\pi}{L_{cr}} z) sin(\frac{\pi}{L_{cr}}
z')}{\Delta\left(1\,-\,i\frac{\delta\omega}
{\Omega}\,+\,\frac{t^2}{2 \tau_{\lambda}^2 \Omega \tau} \right)}
\end{equation}
The further calculations follow the same lines as those in the
case of stationary scatterers.

According to (37) and (38) there exist short-range time correlations
$F_1(t)$ and $$F_3(t)\,\sim\,\frac{\Delta^{3/2}}{1\,+\,\frac{t^2}
{2 \tau_{\lambda}^2 \Omega \tau}}$$ with a scale $\tau_{\lambda}
\sqrt{\Omega \tau} $
and long-range time correlation
$ F_2(t)\,\sim\, \Delta^{3/2} \exp (- \Omega t) $.

\section{Conclusions}
In conclusion, we have calculated angular, frequency and time correlation
functions for the transmission of light through a disordered amplifying
medium.
The correlation function in the case of stationary dielectric function
fluctuations
consists of three terms like
in an elastic scattering
medium.
The angular dependence of the correlation function is exactly the same
as in an elastic medium with tube geometry.
This similarity is due to the absence in the emitted radiation of memory about the incident
radiation in  both cases.
Frequency correlation has a scale
$ \Omega\,=\,2 \Delta/\tau_{amp} \rightarrow 0 $ at
$ \Delta \rightarrow 0$.
It is the same for all three terms.

Interaction between diffusion modes is strongly enhanced near the
lasing threshold.
The enhancement of transmission fluctuations is more pronounced than
the enhancement of weak localization near the threshold.

In the case of nonstationary dielectric function fluctuations, speckle correlation function consists of terms
with short- and long-range time correlations.
The short-range correlation has a scale
$ \tau_{\lambda} \sqrt{ \Omega \tau} \rightarrow 0$ at
$\Delta \rightarrow 0$.
Interaction between diffusion modes is suppressed by the motion
of the scatterers.

Although our calculations were performed in the case of all sample
dimensions being larger than the mean free path, we believe that our
results also apply to the case $L_z\leq l_0$.
Probably only the ''interaction parameter''
$ (g\Delta^2)^{-1}$ and definition of parameter $\Omega$ will be modified.

Experimentally it would be interesting to realize the crossover
between the both cases in one system.
Probably, a good candidate is an amplifying
liquid crystal near the phase transition,
which unduces strong changes of transparency.
Recent experiments \cite{Johnson93} show that it is possible to realize
the multiple scattering regime in a liquid crystal.

\newpage
\section*{Figure captions}

1. Figure 1.

Schematic picture of an amplifier-generator phase diagram for a disordered
slab of thickness $L_z$ with a homogeneous amplification length $l_{amp}$.
In the region below the curve the slab is a random amplifier.
Above the curve it is a random generator.\\
2. Figure 2.

Diagram for the diffusion propagator $P$.
Solid lines correspond to the average Green's function $D$, dashed lines
denote scattering.\\
3. Figure 3.

The diagrammatic representation of $C^{(1)}$.\\
4. Figure 4.

The diagrammatic representation of the correlation function of the Langevin
currents which are proportional to the average intensity.\\
5. Figures 5a,b,c.

The diagrammatic representation of the correlation function of the Langevin
currents which are proportional to the average current.
We denote the average current by a shaded vertex.
\end{document}